\documentclass[]{aa}
\usepackage{graphicx,longtable,lscape}
\usepackage{txfonts,psfig}
\usepackage[]{natbib}
\def \sw {{\it Swift}}
\def\ltsima{$\; \buildrel < \over \sim \;$}
\def\lsim{\lower.5ex\hbox{\ltsima}}
\def\gtsima{$\; \buildrel > \over \sim \;$}
\def\gsim{\lower.5ex\hbox{\gtsima}}

\newcommand{\be}{\begin{equation}}
\newcommand{\en}{\end{equation}}

\begin{document}

  \title{The 54 days orbital period of AX~J1820.5--1434 unveiled by Swift.}
   \author{A.\ Segreto $^{1}$, V.\ La Parola\inst{1},  G.\ Cusumano, 
 A.\ D'A\`i$^{2}$, N.\ Masetti $^{3}$, S.\ Campana$^{4}$}

   \offprints{A. Segreto, segreto@ifc.inaf.it}
   \institute{INAF - Istituto di Astrofisica Spaziale e Fisica Cosmica di Palermo, 
        Via U.\ La Malfa 153, 90146 Palermo, Italy 
\and
 Dipartimento di Fisica e Chimica, Universit\`a di Palermo, via Archirafi 36, 90123, Palermo, Italy 
\and
INAF - Istituto di Astrofisica Spaziale e Fisica Cosmica di Bologna,
via Gobetti 101, 40129, Bologna, Italy
\and
INAF - Brera Astronomical Observatory, via Bianchi 46, 23807, Merate (LC), Italy 
}
%   \date{}
%\date{Received 1 March 2007/Accepted 4 April}

\abstract
{The hard X-ray survey that Swift-BAT has been performing since late 2004 has
provided a considerable database for a large number of sources whose hard X-ray
emission was poorly known.} 
%For many Galactic sources this has allowed the detection of their orbital period.}
{We are exploiting the BAT survey archive to improve the temporal and spectral
characterization of the Galactic hard-X-ray sources. In this letter we focus on
the study of the high mass X-ray binary  AX~J1820.5--1434.}
{All the data relevant to AX~J1820.5--1434 have been extracted from the
BAT survey archive and analyzed using a folding technique to search for
periodical modulations. A broad-band spectral analysis was also performed
complementing the BAT spectrum with the available Swift-XRT and XMM-Newton pointed observations.    }
{A timing analysis has revealed the detection of a coherent signal
at {\bf P$_0=54.0\pm 0.4$ d}, that we interpret as the orbital period of 
the binary system. When folded with a period of P$_0$, the light curve shows an
asymmetric profile, with a minimum roughly consistent with zero intensity.
The broad band spectral analysis performed coupling Swift-XRT, XMM-Newton and 
Swift-BAT spectra confirms that the source emission is well modeled with a 
strongly absorbed power law with no evidence for a high energy exponential cutoff.}
{}
\authorrunning {A.\ Segreto et al.}
\titlerunning {The 54 days orbital period of AX~J1820.5--1434 }

\maketitle

\section{Introduction\label{intro} }

Since November 2004 the Burst Alert Telescope (BAT, \citealp{bat}) on board  Swift 
\citep{swift} has been performing  
a continuous monitoring of the sky in the hard X-ray domain (15--150 keV).
Thanks to its large field of view  (1.4 steradian half coded)
and to its pointing strategy, 
%with several targets pointed within the same day, 
BAT covers a fraction of the sky of  50\% to 80\% every day.
This monitoring  has been very  successful to unveil the binary nature
of many Galactic  sources (e.g. \citealp{corbet1, corbet2, cusumano10,
laparola10, dai11}).

In this Letter we present a temporal and spectral analysis of the Swift data 
collected on the AX~J1820.5--1434, a high mass X-ray binary (HMXB) discovered during the ASCA Galactic plane
survey
%, in a region  between the Galactic center and the Scutum region 
at R.A.$_{J2000}$ = 18h20m29.5s, Dec$_{J2000}$ = $-14^{\circ}34' 24''$ with an error radius
of 0.5' \citep{kinugasa98}. The timing analysis of the ASCA data allowed the detection of a
coherent pulsation  with a period of $152.26 \pm 0.04$ s. 
The spectrum was modeled with a strongly absorbed 
(N$_{\textrm H}=9.8\pm 1.7\times 10^{22}$ cm$^{-2}$) power-law with photon index
$\Gamma= 0.9 \pm 0.2$ and a 6.4 keV iron line with equivalent width $\sim100$ eV.
The average absorbed flux was $2.3\times 10^{-11}$ erg s$^{-1}$ cm$^{-2}$ in the 2--10 
keV energy band.
The source was detected in the hard X-rays with INTEGRAL \citep{atel155}, with an 
average 18--60 keV flux of $\rm \sim 9.3~erg~s^{-1} cm ^{-2}$ between 
March and April 2003, while the 
non-detections in later observations indicated its transient nature
\citep{filippova05}. 
%It is also reported as a transient source in the 
%\citet{bird10} and \citet{krivonos12} INTEGRAL catalogues.
An optical counterpart was proposed by \citet{negueruela07} to be the mid or 
early B-type star USNO-B1.0 0754-0489829. 
However, thanks to a refinement of the X-ray position achieved through two 
later XMM-Newton and Chandra observations (R.A.$_{J2000}$ = 18h20m30.09s, 
Dec$_{J2000}=-14^{\circ}34' 23.52''$ derived with Chandra, with an error 
circle of $0.64''$, \citealp{kaur10}), this counterpart candidate was 
rejected because inconsistent with the new position.  No optical counterpart is visible in the Digitized Sky 
Survey while a bright near infrared (NIR) counterpart was found in the 
Two Micron All Sky Survey (2MASS) at R.A.$_{J2000}$ = 18h20m30.10s, 
Dec$_{J2000}= -14^{\circ}34' 22.9''$ (error radius $0.1''$). 
The J, H and Ks magnitudes of the NIR source are 15.41, 13.25 and 11.75, 
respectively.  The bright NIR counterpart could be either an early-type single 
star or a blend of a few nearby stars not resolved due to poor resolution of 
the 2MASS observations. As a consequence, the actual counterpart of AX~J1820.5--1434 remains 
still uncertain  \citep{kaur10}. 
The spectral parameters derived from the XMM-Newton observation are consistent with 
those reported by Kinugasa et al. (1998), with an absorbed flux a factor of 
$\sim15$ lower. The timing analysis confirmed the  pulse period reported by \citet{kinugasa98}. An 
average spin-period derivative of $(3.00\pm0.14)\times 10^{-9}$ s s$^{-1}$ was 
determined using the previous spin-period measurement. 
%
%This Letter is organized as follows. Sect.~2 describes the Swift data
%reduction. Sect.~3 reports on the timing analysis.
%Sect. 4 describes the spectral analysis.  In Sect.~5 we shortly
%discuss the results.

\begin{figure*}%%%%%%%%%%%%%%%%%%%%%%%%%%%%%%%%%%%%%%%%%%%%%%%%%%%%% PAP VII FIGURE 1
\begin{center}
%\vspace{-1.5truecm}
\centerline{\includegraphics[width=13cm,angle=0]{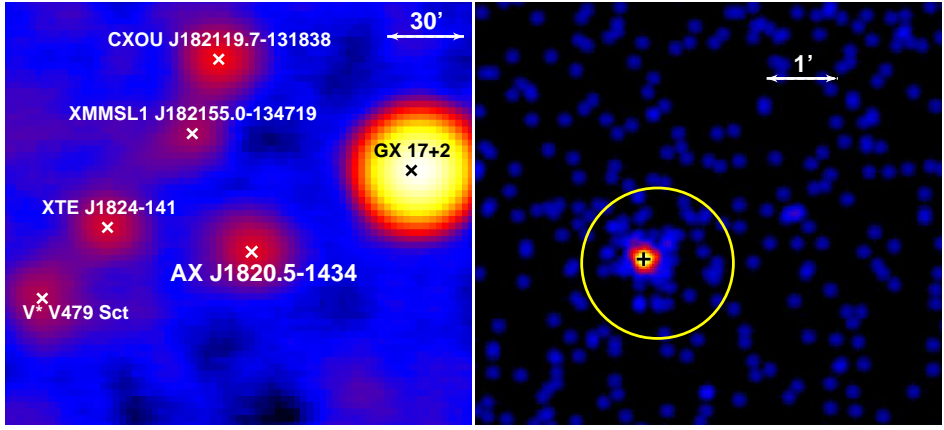}}
\caption[]{{\bf Left panel:} 15--60 keV BAT significance map of the sky region around
AX~J1820.5--1434.
{\bf Right panel:} 0.2--10 keV XRT image with superimposed the BAT error box (1.08 arcmin,
yellow circle) and the position of the NIR counterpart (black cross) suggested by
\citet{kaur10}.
                }
                \label{map}
        \end{center}
\vspace{-0.5truecm}
        \end{figure*}

\section{Observations and data reduction\label{data}}

The Swift-BAT survey data (2004/11 -- 2012/03) retrieved from the HEASARC public 
archive\footnote{http://heasarc.gsfc.nasa.gov/docs/archive.html}
were  processed using the {\sc batimager} code \citep{segreto10}, dedicated to
the processing of coded mask instrument data.
%, that provides all sky maps in
%several energy bands and standard scientific products for the detected sources.
AX~J1820.5--1434 is detected with a significance of  23.6  standard deviations in 
the 15--150 keV all sky map, and of  24.9 standard deviations 
in the 15--60 keV all-sky map, where its signal-to-noise 
is maximized (Fig.~\ref{map}, left panel). The latter energy band was used to extract 
the light curve with the maximum resolution allowed by the Swift-BAT survey data. 
The background subtracted spectrum averaged over the entire survey period was 
extracted in eight energy channels and analyzed using the BAT redistribution 
matrix available in the Swift calibration 
database\footnote{http://swift.gsfc.nasa.gov/docs/heasarc/caldb/swift/}.

\begin{table}
\caption{XRT and XMM-Newton observation log.\label{log}}
\scriptsize
\begin{tabular}{l l l r r r r}
\hline
      & ObsID       & $T_{start}$ & $\Delta$T      &Exp. T  & Orb. phase & Rate\\
      &             &  (MJD)      & (s)            &  (s)    &            & c/s\\  \hline \hline
XRT   &             &             &                &         &            &        \\
1     & 00036168001 &  54155.66   &24404         & 6006  & 0.21       & 0.013 \\
2     & 00037882001 &  55031.10   &  171         &  171  & 0.41       & $<0.081$\\
3     & 00044152001 &  56216.54   &  480         &  480  & 0.35       & 0.16\\ \hline 
XMM   &             &             &                &         &            &     \\
Epn   & 0511010101  &54373.29     & 8150         &  8150 & 0.23     & 0.080 \\
EMOS1 &  --         & --          &10733         & 10733 &    --      & 0.025\\
EMOS2 &  --         & --          &10753         & 10753 &    --      & 0.025\\
\hline
% 4 source 1  back     1.3 sigma    8.88275
\end{tabular}
\tablefoot{$\Delta$T is the observation lenght; Exp. T is
the net exposure time; the orbital phase is referred to the profile in
Figure~\ref{period}b. The count rate reported for the Swift-XRT 
ObsID 00037882001 is a 3$\sigma$ upper limit.}
\vspace{-0.5truecm}
\end{table}

Two Swift-XRT observations were used in this work to perform broad band
spectroscopy: ObsID 
00036168001, performed on 2007 February 24, with CCD substrate voltage set 
at 0V and with a net exposure of 6 
ksec, and ObsID 00044152001, 
performed on 2012 October 16, for $\sim 500$ s of exposure time 
(CCD substrate voltage set at 6V). Both
pointings are in Photon Counting observing mode \citep{hill04}. A third 
observation (ObsID 00037882001, 2009 July 19) has an exposure time of only 
170 s, and the source was not detected.     
The data were processed with standard procedures ({\sc xrtpipeline} v.0.12.4), 
filtering and screening criteria, adopting a 0-12 grade selection.
The source events were extracted from a circular region 
(20 pixel radius, with 1 pixel = 2.36 arcsec) centered on the 
source centroid, determined using the task {\sc xrtcentroid} 
 (R.A.$_{J2000}$ = 18h20m30.1s, Dec$_{J2000}=-14^{\circ}34' 23.8''$, with 90\% confidence region of 
 $3.98''$ radius).  
Figure~\ref{map} (right panel) shows the XRT image with the position of the NIR 
counterpart, that is offset with respect to the XRT source of $0.9''$.  
The background for the spectral analysis was extracted from an annular
region  with inner and outer radii 30 and 70 pixels, respectively. XRT ancillary 
response file were generated with
{\sc xrtmkarf}\footnote{http://heasarc.gsfc.nasa.gov/ftools/caldb/help/xrtmkarf.html};
we used the spectral redistribution matrix v011 for ObsID 00036168001 and v013 for 
ObsID 00044152001. The spectral analysis was 
performed using {\sc xspec} v.12.5.
The event arrival times were corrected to the SSB  using the task {\sc
barycorr}\footnote{http://http://heasarc.gsfc.nasa.gov/ftools/caldb/help/barycorr.html}.

We also extracted the XMM-Newton spectrum of the source, from the same
pointings whose analysis was 
performed in \citet{kaur10}. We  extracted both Epic-PN and MOS spectra, using standard 
pipelines of SAS tools. We selected a circular region of 10 arcsec radius centered on the 
source position for source spectra and a similar region from a nearby source-free area for 
the background spectra. We checked the consistency of our analysis by  successfully 
reproducing the best-fitting  model of \citet{kaur10}.

\begin{figure}
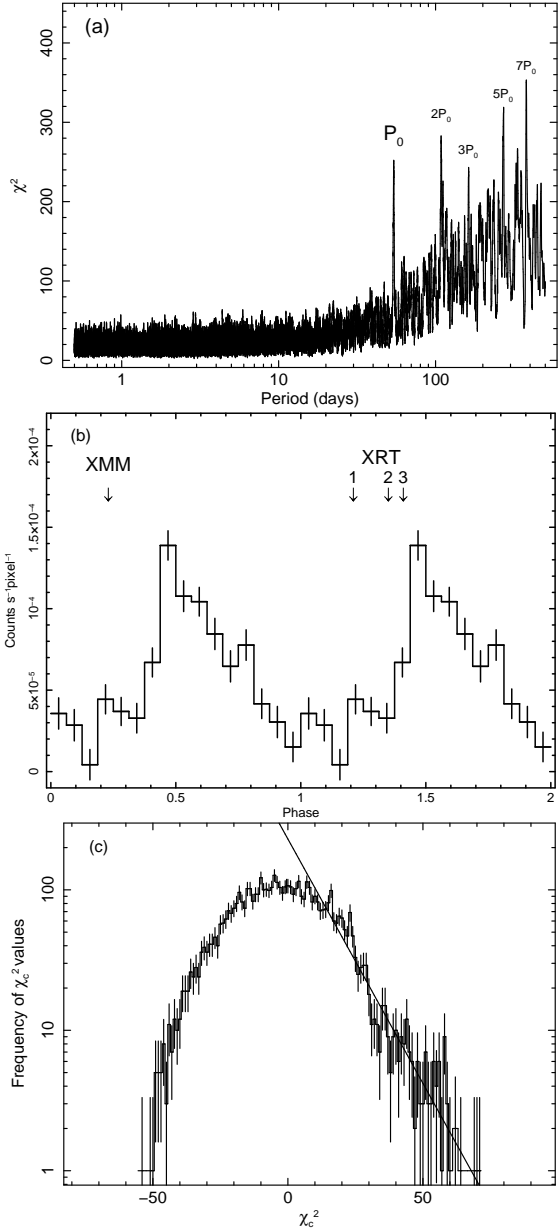
%%%%%%%%%%%%%%%%%%%%%%%%%%%%%%%%%%%%%%%%%%%%%%%%%%%%% PAP VII FIGURE 2
\begin{center}
%\vspace{-1.5truecm}
\centerline{\includegraphics[width=5.4cm,angle=270]{figura2.ps}}
\centerline{\includegraphics[width=5.4cm,angle=270]{figura4_new.ps}}
\centerline{\includegraphics[width=5.4cm,angle=270]{figura_histo.ps}}

%\vspace{-2.5truecm}
\caption[]{{\bf a}: Periodogram of \sw-BAT (15--60\,keV) data for 
AX~J1820.5--1434. 
{\bf b}: Light curve folded at a period {\bf P$_0= 54.0$\,d}, with 16 phase 
bins. The arrows mark the orbital phase of the Swift-XRT and XMM-Newton observations.
{\bf c}: Histogram distribution of the $\chi^2_{c}$ values; the solid line is
the exponential function that best fits the right tail of the distribution.}      
         
                \label{period} 
        \end{center}
\vspace{-0.5truecm}
        \end{figure}

        %%%%%%%%%%%%%%%%%%%%%%%%%%%%%%%%%%%%%%%%%%%%%%%%%%%%%%%%%
        \section{Timing analysis\label{sfxt7:timing}}
        %%%%%%%%%%%%%%%%%%%%%%%%%%%%%%%%%%%%%%%%%%%%%%%%%%%%%%%%%

A folding timing analysis \citep{leahy83} was applied to the 15--60 keV Swift-BAT light curve
searching for the presence of periodic modulations with period between 
 0.5 and 500\,d. 
{\bf The method consists in building a light curve profile at
different trial periods by folding the photon arrival times in $N$ phase bins. 
For each resulting light curve  the $\chi^2$ with respect to
the average count rate is evaluated:
a periodic pulsation corresponds to a large value of $\chi^2$.} 
The resolution of the period search is $\Delta P=P^{2}/(N \,\Delta T)$,
where P is the trial period, $N=16$ is the number of phase bins used to build the trial
profile, and $\Delta T=$228  Ms is the data time span. 
The profile for each trial period was built weighting the rates 
by the inverse square of the corresponding statistical error (see \citealp{cusumano10}). 
This allows to cope with the large spread of statistical errors that
characterizes the Swift-BAT survey data, mainly due to the wide range of off-axis 
directions in which the source is observed. 
The resulting periodogram (Figure~\ref{period}a) shows  
several features emerging over the noise: the one with the shortest period 
is at {\bf P$_0=54.0\pm0.4$ d} ($\chi^2\sim254$), where P$_0$ and its 
error are the centroid and the standard deviation obtained modeling this 
feature with a Gaussian function, while the 
other features result to be multiples of P$_0$.
The intensity profile (Fig.~\ref{period}b) obtained by folding the data at P$_0$ with T$_{\rm
epoch}$=54684.809 MJD shows a large asymmetric single peak profile with a 
minimum which is marginally consistent with zero emission.
For larger values of the trial period the periodogram shows the typical 
average $\chi^2$ increase which is expected when the source emission is 
characterized by long term time variability (red noise). In presence of red
noise the $\chi^2$ statistics  cannot  be applied and the 
significance of the feature is to be evaluated with respect
to the local noise fluctuation of the periodogram. 
Thus we fit the periodogram with a second order polynomial
and subtract the best fit trend from the  $\chi^2$ values, obtaining a
``corrected'' periodogram $\chi^2_{c}$. The value of $\chi^2_{c}$ at  P$_0$
is $\sim194$.

In Fig~\ref{period}c we show the histogram built from the 
corrected periodogram selecting the values 
in the period interval between 24 and 104 {\bf d } (as this interval is
characterized by a noise level quite consistent with the noise level at P$_0$) 
and excluding those falling 
in an interval of $\pm\Delta P_0$ around P$_0$.
We then use an exponential function to extrapolate the histogram right tail and compute its
area for $\chi^2_{c}>\chi^2_{c}(P_0)=194$. 
Normalizing this value to the total area of the histogram we obtain $1.0\times
10^{-7}$ that represents the
probability of random occurrence for $\chi^2_{c} > 194$. 
This corresponds to a significance of 5.4 standard deviations in
Gaussian statistics.

\begin{figure*}%%%%%%%%%%%%%%%%%%%%%%%%%%%%%%%%%%%%%%%%%%%%%%%%%%%%% PAP VII FIGURE 1
\begin{center}
%\vspace{-1.5truecm}
\centerline{\includegraphics[width=20.0cm]{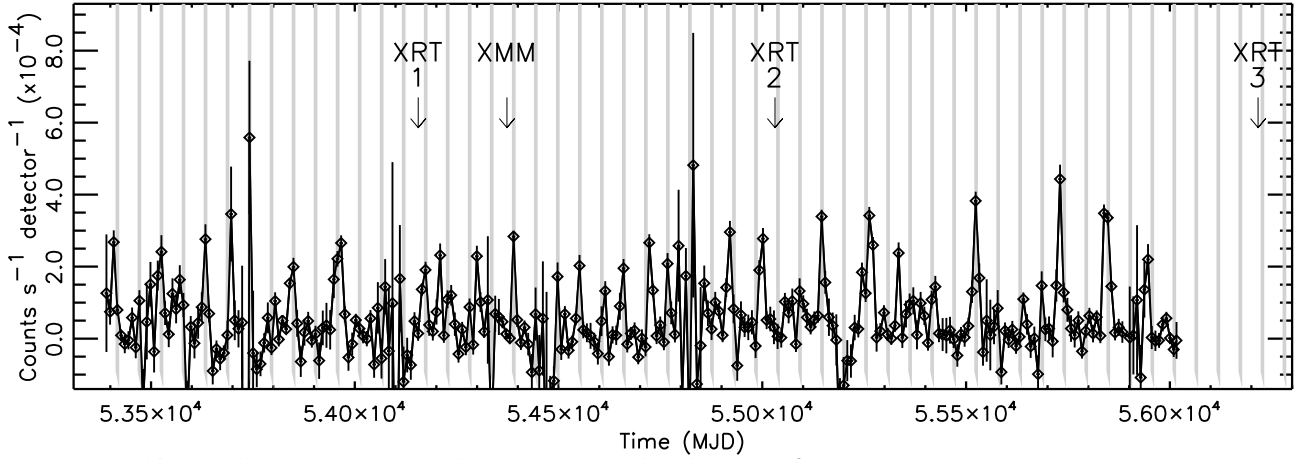}}
\caption[AX~J1820.5--1434 Swift-BAT light curve]{15--60 keV Swift-BAT light
curve.  Each point represents a time interval {\bf of  P$_0/6=9.0$ d.}
The vertical shaded bars mark the phase interval 0.44--0.62  of the 
folded profile in Figure~\ref{period}b.
The arrows mark the epoch of the Swift-XRT and XMM-Newton observations.
The third Swift-XRT observation is later than the 88-month BAT survey data 
analised in this paper.
                }
                \label{lc}
        \end{center}
\vspace{-0.5truecm}
        \end{figure*}

\noindent Figure~\ref{lc} shows the 15--60 keV light curve of 
AX~J1820.5--1434 with a bin time {\bf of P$_0/6=9.0$ d}. The vertical shaded bars
mark the phase interval 0.44--0.62  of Figure~\ref{period}b. In $\sim 50\%$ of 
the cycles we observe a significant intensity enhancement in this phase
interval. The source shows also a few episodes of significant intensity
enhancements at different phases.
The Swift-XRT observations correspond to a phase interval between 0.2 and 0.4 in
Figure~\ref{period}b. The variations of the rate observed among the observations
(Table~\ref{log}) are in fair agreement with the shape of the Swift-BAT folded profile
with the highest rate observation (ObsID 00044152001) corresponding to the profile peak.
The statistic content of the Swift-XRT observations is too low ($\sim 70$ counts in
observations 1 and 3) to see the modulation of $\sim 152$ s reported by
\citet{kinugasa98} and confirmed by \citet{kaur10} with the XMM-Newton data.

%%%%%%%%%%%%%%%%%%%%%%%%%%%%%%%%%%%%%%%%%%%%%%%%%%%%%%%%%
\section{Spectral analysis\label{sfxt7:xrt}}
%%%%%%%%%%%%%%%%%%%%%%%%%%%%%%%%%%%%%%%%%%%%%%%%%%%%%%%%% 

Before performing a broad band spectral analysis we checked the range of 
spectral variations observed in the available data sets.
The spectra of the two Swift-XRT pointed observations where AX~J1820.5--1434 is 
detected (see Sect. 2) and of the XMM observation were fitted simultaneously with an absorbed power law 
and constraining the absorption column density and the photon index to be the 
same for both data set. The residuals resulting from the best fit model were 
consistent with each other within the errors.  
The absorption column density  was $\rm 11.3^{+2}_{-2} \times 10^{22} cm^{-2}$ 
while the photon index was $1.6^{+0.3}_{-0.3}$. 
The same check was done on 
the Swift-BAT spectra produced dividing the BAT data into four 22-month 
long time intervals and into three different orbit phase intervals 
(0.375-0.4375 plus 0.625--0.875, 0.4375--0.625 and 0.875--1.375, see 
Figure~\ref{period}b), that were fitted with 
a power law. As for the Swift-XRT spectra, the photon index was tied to be the same in the 
three datasets. 
The best fit residuals show the same trend for all the BAT datasets with a best fit photon index 
of $2.6^{+0.1}_{-0.1}$.

\begin{figure}%%%%%%%%%%%%%%%%%%%%%%%%%%%%%%%%%%%%%%%%%%%%%%%%%%%%% PAP VII FIGURE 2
\begin{center}
\centerline{\includegraphics[width=5.1cm,angle=270]{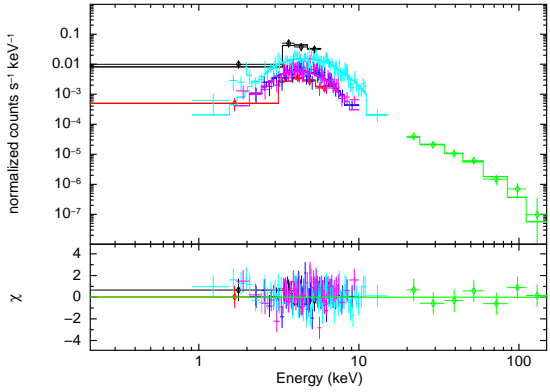}}
\caption[]{ AX~J1820.5--1434
  broad band (0.2-150 keV ) spectrum. {\bf Top panel}:
XRT (black and red data), EPIC-pn (cyan data), EPIC-MOS1 and EPIC-MOS2 (magenta and blue
data), BAT data  (green data) and best fit {\tt phabs*(powerlaw)} model.
The Swift data are marked with diamonds.
{\bf Bottom panel}: Residuals in unit of standard deviations.
}
                \label{spec}
        \end{center}
\vspace{-0.5truecm}
        \end{figure}

The broad band spectral analysis was therefore performed coupling the 15-150 keV 
BAT spectrum averaged over 88 months, the two Swift-XRT spectra and the XMM-Newton spectra. 
In order to take into account flux variations and intercalibration issues, 
we included in the model a multiplicative normalization factor, fixed to 1 for the
Swift-XRT spectrum of Obs ID 00044151001 and left free for all the other spectra.
A strongly absorbed ($\rm N_H\sim 1.7\times 10^{23} cm^{-2}$) power law 
was {\bf adequate} to describe the data with an acceptable 
$\chi^2$ of  158.4 (182 d.o.f.).
Figure~\ref{spec} shows data, best fit model  and residuals. 
Table~\ref{fit} reports the best fit parameters               
(quoted errors are given at 90\% confidence level for a single parameter).
We also tried a model consisting in  an absorbed power-law with a high energy exponential cutoff
({\tt phabs*cutoffpl}) without any significant fit improvement ($\chi^2$=145.4 with 181 d.o.f.).

\begin{table}
\caption{Best fit spectral parameters. \label{fit}}
\scriptsize
\begin{tabular}{ r l l}
\hline
Parameter       & Power law  & Units    \\ \hline \hline 
$\rm n_H$       &$1.69^{+0.17}_{-0.15} \times 10^{23}$ & $\rm cm^{-2}$\\
$\Gamma$        &$2.44^{+0.15}_{-0.14}$&          \\
$N$             &$4.6^{+2.0}_{-1.4}\times 10^{-2}$ &$\rm ph/(keV cm^{2} s)$ @ 1 keV  \\
$\rm C_{V0}$    &$0.09 ^{+0.04}_{-0.03}$&\\
$\rm C_{BAT}$   &$0.8 ^{+0.04}_{-0.03}$&\\
$\rm C_{EPIC}$  &$0.06 ^{+0.02}_{-0.01}$&\\
$F_{0.2-10~keV}$&$2.2^{+0.4}_{-0.3}\times 10^{-10}$ & $\rm erg~s^{-1}~cm^{-2}$\\
$F_{15-150~keV}$&$4.1^{+0.05}_{-1.25}\times 10^{-11}$ & $\rm erg~s^{-1}~cm^{-2}$ \\
$\chi^2$        &158.4 (182) & \\ \hline
\end{tabular}
\tablefoot{$\rm C_{V0}$, $\rm C_{EPIC}$ and $\rm C_{BAT}$
are the constant factors to be multiplied to the model to match the Swift-XRT data collected with CCD
substrate voltage set to 0 Volt, EPIC data and BAT data, respectively. The constant factors
relevant to the three EPIC detectors set to the same best fit value for the three datasets. We report
unabsorbed fluxes for the characteristic Swift-XRT (0.2--10 keV) and BAT (10--150 keV)
energy bands.}
\vspace{-0.7truecm}
\end{table}

\section{Conclusions\label{conclusion}}
\begin{figure}%%%%%%%%%%%%%%%%%%%%%%%%%%%%%%%%%%%%%%%%%%%%%%%%%%%%% PAP VII FIGURE 1
\begin{center}
%\vspace{-1.5truecm}
\centerline{\includegraphics[width=7cm,angle=0]{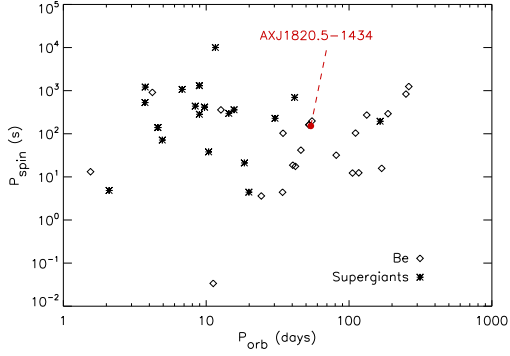}}
\caption[Corbet diagram]{
Corbet diagram for HMXBs with known spin and orbital period. Diamond and star points
represent the Be and supergiant systems, respectively. The red filled circle
marks the position of AX~J1820.5--1434.
                }
                \label{corbet}
        \end{center}
\vspace{-0.5truecm}
        \end{figure}

The entire dataset of Swift and XMM-Newton X-ray data on AX~J1820.5--1434 was analyzed to
explore the temporal and spectral properties of the source. 
The timing analysis performed on the 88-month BAT survey data unveiled a periodic intensity 
modulation at {\bf P$_0=54.0\pm0.4$ d}, with a significance higher than 5.4 standard deviations in
Gaussian statistics. Interpreting this long periodicity as the orbital period of
the binary system, and knowing the spin period \citep{kinugasa98} of the compact star, we have 
located the source position on the Corbet diagram \citep{corbet86}. 
The source lays in the plot region populated by Be/X-ray binaries 
(Fig.~\ref{corbet}). The BAT light curve with a time bin of P$_0/6$ shows 
periodic intensity enhancements mostly concentrated in a narrow phase
interval ($\rm \lesssim P/6$). However, some of them have a longer time
duration, and together with a few enhancement episodes that seems to be not 
related to the same phase interval, this explains the asymmetric 
triangular peak spanning about 50\% of the orbital period in the BAT light 
curve folded at P$_0$. 
This suggests that the accretion onto the
compact object is not strictly related to the periastron passage, as typical 
for highly eccentric Be/X ray binaries. 
The lack of information on the companion star type prevents us from further
investigation on the orbital parameters of the system.
If the counterpart of AX J1820.5$-$1434 proposed 
by \citet{kaur10} is not a blend of objects but rather a single 
source, we can consider its NIR magnitudes and assume the intrinsic 
colors typical of late O/early B-type stars \citep{wegner94}: we find a 
color excess $E(J-K) \sim$ 3.5 mag, that implies a reddening 
$A_V \sim$21 mag using the Milky Way extinction law of \citet{cardelli89} 
and assuming the total-to-selective extinction ratio of \citet{rieke85}. 
Although it is known that the $N_{\rm H}$ can vary on 
short scales throughout the Galactic Plane, and the value of the 
gas-to-dust ratio is not universal in the Galaxy, we can nevertheless use 
this approach to get an (admittedly rough) estimate of the Galactic 
line-of-sight hydrogen column density towards AX J1820.5$-$1434. 
This large extinction naturally explains the lack of detection 
of an optical counterpart for this source and, from the formula 
of \citet{predehl95}, it implies a column density of 
$\sim3.8 \times$10$^{22}$ cm$^{-2}$, substantially lower than that 
inferred from our X--ray spectral analysis and again pointing 
to the presence of further absorbing material local to the system.
Assuming this amount of reddening along the AX J1820.5$-$1434
line of sight and a B0 spectral type for the companion star in this
system, we can infer its distance depending on the luminosity
class of the star. We consider the main sequence and giant cases
(luminosity classes V and III, respectively), as the location of 
the source in the Corbet diagram tends to rule out the presence of 
an early supergiant in this system. From the tabulated absolute 
magnitudes for this type of stars \citep{lang92} we find distances of
$\sim$3.5 and $\sim$6.0 kpc for the main sequence and the giant
companion case, respectively. This range of distance implies a 0.2--10 keV (15--150
keV) luminosity range of $3.2 \times 10^{35} - 9.5 \times 10^{35}$ erg s$^{-1}$ 
($6.0 \times 10^{34} - 1.8 \times 10^{35}$ erg s$^{-1}$).

The set of XMM-Newton, Swift-XRT and Swift-BAT data allows for a broad 
band spectral analysis in the 0.2--150 keV energy band. The data can be 
described with a strongly absorbed ($\rm N_H\sim1.7\times 10^{23}~cm^{-2}$)
power law  with photon index $\Gamma\simeq2.44$. 
We have modified the power law by a high energy cutoff, but the cutoff 
energy we obtain is poorly constrained and the statistical improvement 
is not significant. 

\begin{acknowledgements}
We thank the referee R. Walter whose comments helped to improve the paper.
This work has been supported by ASI grant I/011/07/0.
\end{acknowledgements}

%%%%%%%%%%%%%%%%%%%%%%%%%%%%%%%%%%%%%%%%%%%%%%%%%%%%%%%%%%%%%%%%%%%%%%%%% BIBLIO 
\bibliographystyle{aa}

{}

\end{document}